# TeraFET terahertz detectors with spatially non-uniform gate capacitances

Yuhui Zhang and Michael S. Shur [a)]

*Department of Electrical, Computer and Systems Engineering, Rensselaer Polytechnic Institute, Troy, New York 12180 USA*

[a)] Author to whom correspondences should be addressed. E-mail: shurm@rpi.edu

**ABSTRACT**

A non-uniform capacitance profile in the channel of a THz field-effect transistor (TeraFET) could significantly improve the THz detection performance. The analytical solutions and simulations of the hydrodynamic equations for the exponentially varying capacitance versus distance showed ~10% increase in the responsivity for the 130 nm Si TeraFETs in good agreement with numerical simulations. Using the numerical solutions of the hydrodynamic equations, we compared three different $C_g$ configurations (exponential, linear and sawtooth). The simulations showed that the sawtooth configuration provides the largest response tunability. We also compared the effects of the non-uniform capacitance profiles for Si, III-V, and p-diamond TeraFETs. The results confirmed a great potential of p-diamond for THz applications. Varying the threshold voltage across the channel could have an effect similar to that of varying the gate-to-channel capacitance. The physics behind the demonstrated improvement in THz detection performance is related to breaking the channel symmetry by device geometry of composition asymmetry.

The generation and detection of terahertz (THz) radiation has been a very active area over the recent decades[1-4]. Occupying the frequency range of 100 GHz-30 THz, THz radiation fills the frequency gap between the electronics and photonics. Initially, THz science and technology were explored in the field of astronomy[1,5], and expanded rapidly since 1990s following the development of THz spectroscopy[6,7]. By now, the real-life applications of THz-based technology can be found in various fields, including the biomedical engineering[6,8], VLSI testing[9], wireless communication[10-12], and object sensing[13,14].

A high-quality THz signal detector plays a crucial role in almost all THz applications. The present-day THz sensors are mostly semiconductor-based, with Schottky devices being the dominant product[1,15]. Recently, plasma-wave field-effect transistor (FET) detectors have attracted more attention as they offer high responsivity, ultra-fast response time, and are highly tunable by gate, doping, and channel structures[16-19]. Moreover, plasma-wave THz FETs (TeraFETs) are particularly promising for next-generation communication applications, since these devices exhibited exceptional detection performance in the 200-400 GHz band allocated for beyond-5G communication[10,11,20].

Despite all those merits, plasmonic TeraFETs still face a variety of challenges, including the fabrication burden, requirement for high sensitivity, and noise issues[18,21]. Among others, the detection sensitivity/responsivity is always a key issue. Due to the existence of scattering, the response voltage of TeraFETs can be limited, which impairs the device performance[22-24]. As was discussed in [25], further improvement in the TeraFET responsivity is required to enable their application in 6G communication. In this paper, we show that the implementation of a spatially non-uniform gate capacitance ($C_g$) profile could significantly enhance the responsivity. A non-uniform $C_g$ profile can be realized by altering the barrier layer width. For example, consider a Si FET with a composite barrier layer of $SiO_2$ and $HfO_2$[26-28]. We may profile the $HfO_2$ layer by adjusting the etch rate[29] to achieve a non-uniform $C_g(x)$ (see Section A of Supplementary Material).

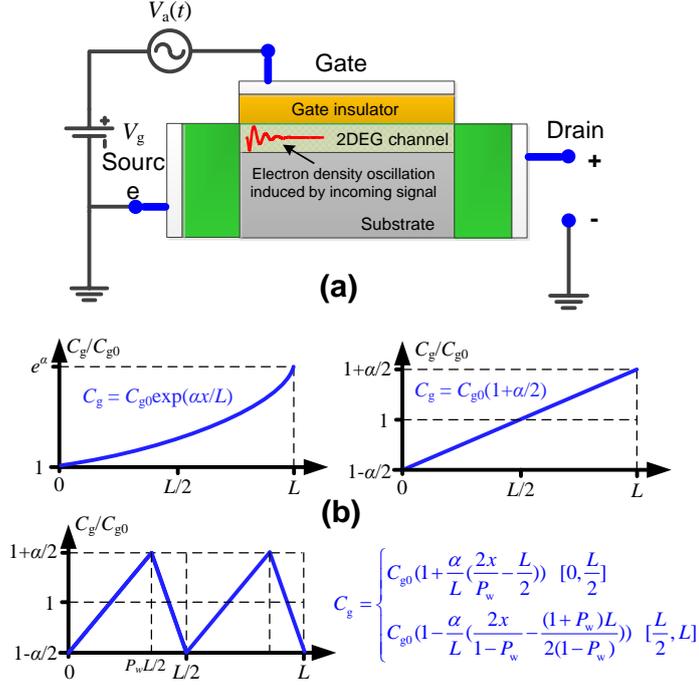

**Fig. 1.** (a) Schematic of THz detection by a TeraFET. $V_a(t)$ represents the radiation-induced AC small signal, and $V_g$ is the DC gate voltage. (b) illustrations of 3 coordinate-dependent gate capacitance ($C_g$) configurations. $C_{g0}$ is the initial constant gate capacitance, α is a modulation factor indicating the total variation of $C_g$ along the channel, $P_w$ is the ratio of rising edge period in one section, $L$ is the channel length. The sawtooth $C_g$ configuration in (b) contains two segments ($N_s$=2). In the simulations of this work, $N_s$ is variable.

Under the DC operation, the gate voltage ($V_g$) and $C_g$ determine the 2D carrier concentration ($n$). Following the unified charge control model (UCCM), the relation between $n$ and $V_g$ can be approximated as[30-32]:

$$n(U) = \frac{C_g \eta V_t}{e} \ln(1+\exp(\frac{U}{\eta V_t})) \quad (1)$$

where $V_t = k_B T/e$ is the thermal voltage ($k_B$: Boltzmann constant, $T$: temperature, fixed at 300 K). $\eta$ is a subthreshold ideality factor. $U = V_g - V_{th} - \varphi$ is the gate-to-channel potential. $V_{th}$ is the threshold voltage, and $\varphi$ is the channel potential. Well above threshold ($U \gg V_t$), UCCM reduces to $n = CU/e$. Given a spatially non-uniform $C_g$, the steady-state density $n_0$ also becomes coordinate-dependent, creating a carrier gradient along the channel at zero drain bias. The existence of density gradient changes the transport of carriers, thus affecting the detection performance of TeraFETs. When we only alter $C_g$, $V_{th}$ also changes. A non-uniform $V_{th}$ can also modify the device characteristics, speed[33], and THz detection performance[34,35]. For now we assume that $V_{th}$ is fixed under a varying $C_g$. The constant $V_{th}$ can be achieved by adjusting the doping level in the device channel[36,37].

The boundary conditions beneficial for a plasma instability are the fixed source voltage of the TeraFET and the drain is fed a current source[38-40]. The Dyakonov-Shur instability occurs due to the difference in the reflection coefficients of the plasma wave from the drain and source sides of the channel. The resulting amplification factor is $(S_e+V_0)/(S_e-V_0)$, where $S_e$ is the plasma wave velocity and $V_0$ is the DC drift velocity. This leads to the generation of plasma instability and the THz radiation[38,41,42]. A non-uniform $C_g$ profile affects the plasma velocity as well as the local Mach number ($M = V_0/S_e$), leading to changes in the amplification factor and the instability growth rate[35,40]. We can therefore adjust the $C_g(x)$ profile to elevate the emitted THz power from TeraFETs and expand the dynamic range of instability regime.

Under the THz detection regime illustrated by Fig. 1(a), the source of the TeraFET is grounded while the drain is open. The plasma wave is initiated at the source and travels towards the drain. The wave gets reflected at the drain without amplification if no DC current source is connected[43]. Due to the nonlinearity of the resonant (or overdamped plasma oscillations) and the boundary conditions breaking symmetry, the AC response is rectified, inducing a DC voltage at the drain which is proportional to the intensity of the impinging THz radiation (at relatively low intensities). Both the DC response voltage and the plasma instability originate from the carrier density oscillations, and their amplitudes are limited by the nonlinearity of the hydrodynamic system[44]. For the plasma instability operation, the transport and amplitude of plasma wave are affected by the non-uniform $C_g$ structure[40,45]. Therefore, we expect that a non-uniform $C_g$ could also alter the continuous-wave THz detection performance of TeraFETs.

To evaluate in more details the effects of varying $C_g$ on the detection performance, we solve the hydrodynamic equations[30,46]:

$$\frac{\partial n}{\partial t} + \nabla \cdot (n\boldsymbol{v}) = 0 \quad (2)$$

$$\frac{\partial \boldsymbol{v}}{\partial t} + (\boldsymbol{v}\cdot\nabla)\boldsymbol{v} + \frac{e}{m}\nabla U + \gamma \boldsymbol{v} = 0 \quad (3)$$

Here, $v$ and $m$ represent the hydrodynamic velocity and effective mass of the carrier, respectively. $\gamma = 1/\tau$ is the momentum relaxation rate, $\tau = 1/\gamma = \mu m/e$ is the momentum relaxation time, and $\mu$ is the field-effect mobility. The energy transport equation is ignored as we fix the operation temperature at 300 K and assume thermal equilibrium. The boundary conditions presented in Fig. 1(a), are $U(0,t) = U_0 + V_a$ and $J(L,t) = 0$, where $U_0 = V_g - V_{th}$ is the DC gate-to-source voltage above threshold, $L$ is the channel length. $V_a = V_{am}\sin(\omega t)$ is an AC small signal voltage generated by the THz radiation shining onto the device.

In the sub-THz range and at room temperature, the TeraFETs typically operate in the non-resonant regime ($\omega\tau \ll 1$). Under this circumstance, $\partial V/\partial t \ll \gamma v$, and thus Eq. (3) reduces to[47,48]

$$\boldsymbol{v} = -\frac{e\tau}{m}\nabla U = -\mu\nabla U \quad (4)$$

Solving (4) together with (1) and (2) assuming a 1-D geometry, and recognizing that $C_g(x)$ is spatially non-uniform, we get:

$$\frac{\partial n}{\partial t} - S_e^2 \tau\left(\frac{\partial^2 n}{\partial x^2} - \frac{\partial(\alpha_1 n)}{\partial x}\right) = 0 \quad (5)$$

where $\alpha_1 = (1/C_g)(\partial C_g/\partial t)$ is a gate capacitance modulation factor, $S_e$ is the plasma wave velocity given by $S_e = \sqrt{\frac{\eta e V_t}{m}(1+\exp(-\frac{U}{\eta V_t}))\ln(1+\exp(\frac{U}{\eta V_t}))}$. For the small signal operation, one can assume $S_e(U) = S_e(U_0)$.

Inspecting Eq. (5), we can see that the non-uniform $C_g$ affects the carrier distribution via $\alpha_1$ and the term $\partial(\alpha_1 n)/\partial x$. For uniform $C_g$, $\partial(\alpha_1 n)/\partial x = 0$ and Eq. (5) reduces to Eq. (10) in [47]. Note that both $\alpha_1$ and $\partial\alpha_1/\partial x$ are functions of $x$, given an arbitrary $C_g(x)$ profile, the solution of $n$ can be highly nonlinear. To reduce the analytical burden, we first consider a constant $\alpha_1$ (i.e. $C_g(x)$ is an exponential function of $x$). With this simplification, Eq. (5) can be solved together with the boundary conditions. We convert the voltage boundary conditions into their density counterparts, and expand with respect to $V_{am}$[47]:

$$\begin{cases} n(0,t) \approx \begin{cases} \frac{C_g \eta V_t}{e}\exp(\frac{U_0}{\eta V_t})(1+\frac{V_{am}\cos\omega t}{\eta V_t}+\frac{V_{am}^2}{4(\eta V_t)^2}) & \text{(sub-}V_{th}\text{)} \\ \frac{C_g(U_0+V_{am}\cos\omega t)}{e} & \text{(above }V_{th}\text{)} \end{cases} \\ \left(\frac{\partial n}{\partial x} - \alpha_1 n\right)\bigg|_{x=L} = 0 \end{cases} \quad (6)$$

Solving Eq. (5) together with (6) under the subthreshold condition, averaging the solution of $n$ over time, and comparing with previous response solutions[24], we obtain a semi-empirical solution for DC response voltage ($dU$, the DC source-to-drain voltage) (see Section B of Supplementary Material for detailed derivations):

$$dU = \frac{eV_{am}^2}{4mS_e^2}\left(1+\beta-\frac{1+\beta\cos(2k_r L)}{\cosh(k_1 L)\cosh(k_2 L)}\right) \quad (7)$$

where

$$k_1 = \frac{\alpha_1}{2} + \sqrt{\left(\frac{\alpha_1}{2}\right)^2 + ik_0^2}$$
$$k_2 = \frac{\alpha_1}{2} + \sqrt{\left(\frac{\alpha_1}{2}\right)^2 - ik_0^2} \quad (8)$$

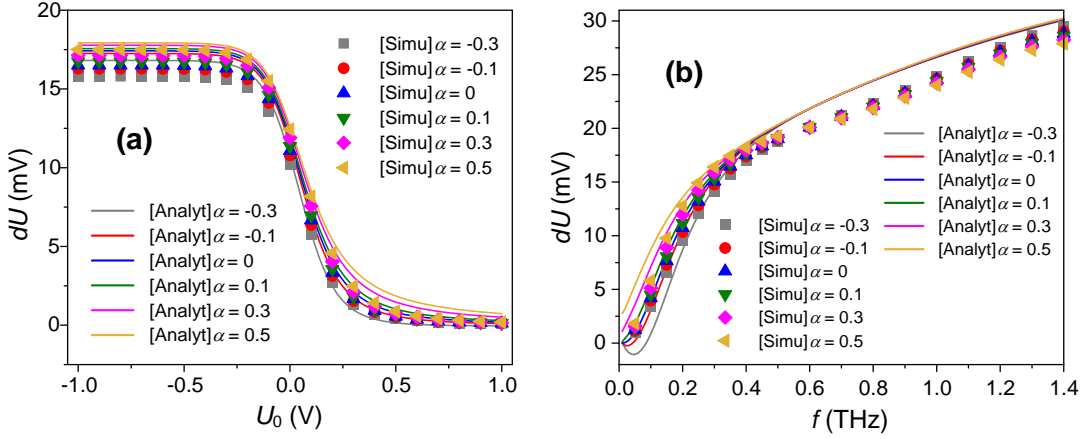

**Fig. 2.** (a) DC response voltage ($dU$) as a function of DC gate bias ($U_0 = V_{g0} - V_{th}$) under the radiation frequency $f$ = 350 GHz and (b) $dU$ as a function of $f$ under $U_0$ = -0.4 V. Si TeraFETs, parameters: $m/m_0$ = 0.19, $\mu$ = 0.1 m$^2$/Vs, $L$ = 130 nm. $C_g = C_{g0}\exp(\alpha x/L)$, with $\alpha$ = -0.3~0.5. The solid lines represent analytical curves, while scatters are simulation data points.

are complex wave numbers, $\beta = 1/\sqrt{1+(\omega\tau)^{-2}}$.

$$k_r = \frac{\alpha_1}{2} + \sqrt{\frac{\sqrt{(\frac{\alpha_1}{2})^4 + k_0^4} + (\frac{\alpha_1}{2})^2}{2}}$$

is the real part of $k_1$ or $k_2$. $k_0 = \sqrt{\omega/S_e^2\tau}$ is the amplitude of $k$ in uniform $C_g$ channels.

Eq. (7) has a similar form as the expression of $dU$ under a uniform $C_g$ (see Eq. (24) in [24]). The main difference lies in the dispersion relation and wave numbers. This indicates that a non-uniform $C_g$ changes the transport and decay features of collective carrier waves in the TeraFETs. As $\alpha_1$ increases, $k_r$ also rises, suggesting that the channel can now accommodate more full waves (or a larger total phase of wave(s)). With more "ups and downs" in the channels, the DC component could develop more efficiently.

To validate the theory proposed above, we compare the theory results with numerical simulations. The model used for simulation is the same as the one in [22], except that we assume thermal equilibrium here and ignore the energy relaxation equation. This model has been validated with the experimental data and proved to be effective[22,43,49]. Fig. 2(a) shows the simulated and analytical $dU$ as a function of $U_0$ for Si TeraFETs under $f$ = 350 GHz. The spatial distribution of $C_g$ is designed as $C_g = C_{g0}\exp(\alpha x/L)$, where $C_{g0}$ is a constant capacitance, $\alpha$ is the $C_g$ modulation factor. Under this configuration, $\alpha_1 = \alpha/L$. As seen in Fig. 2(a), the simulated $dU$ data show the same variation trend as that of analytical curves, i.e. the response increases with the decrease of $U_0$, and saturates in the subthreshold region. It is worth noting that this $U_0$ dependence profile results from the infinite load resistance (ideal open drain) and zero gate leakage current. For a finite load resistance or a non-zero leakage current, the response will be attenuated in the deep subthreshold region due to the dramatic increase of the channel resistance (see Section C of Supplementary Material for more details). Moreover, both simulation and analytical results show that $dU$ increases with rising $\alpha$ in the subthreshold region. Beyond threshold, the $\alpha$-dependence of $dU$ weakens as $U_0$ increases. However, the effect of a non-uniform profile is significant near the threshold, where most TeraFET THz detector operate.

Fig. 2(b) shows the frequency ($f$) dependence of $dU$ under different $\alpha$ values at $U_0$ = -0.4 V. A good qualitative (even quantitative) agreement between the simulation and analytical results is observed at $f$ < 400 GHz ($\omega\tau\ll 1$, deep non-resonant region). The best agreement is achieved within 200-400 GHz band, overlapping the frequency range of beyond-5G communication. For $f$ > 500 GHz, the Si device approaches the resonant regime, and thus the analytical curves deviate from the simulation data. But the general variation trend remains the same. In the non-resonant region, a larger $\alpha$ leads to a larger $dU$. While for $f > 500$ GHz, the $\alpha$ dependence of $dU$ gradually reverses with the increase of frequency, as presented in Fig. 2(b) by the simulation data. In the analytical theory, if we change the signs of square root terms in $k_1$ and $k_2$ (corresponding to another pair of solution of the dispersion equation), the $\alpha$ dependence of $dU$ also reverses. Therefore, we may attribute the reversal of $\alpha$ dependence to the changes in wave propagation properties as the device enters the resonant mode. The results in Fig. 2 demonstrate that we successfully manipulated the DC response of a Si FET by an exponential $C_g$ structure under subthreshold, non-resonant condition, and the proposed analytical theory shows a fair agreement with the simulation results.

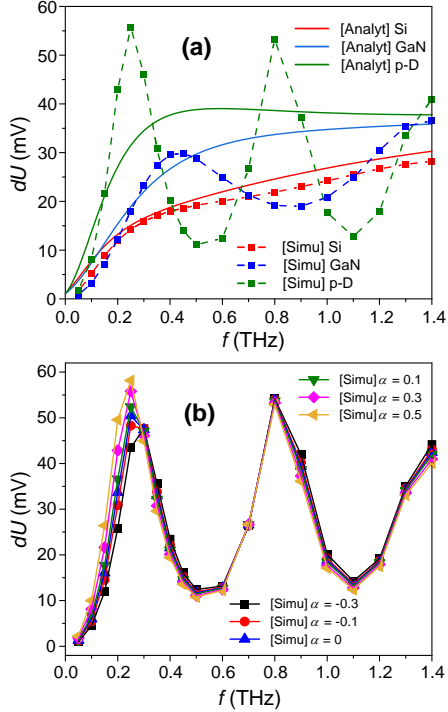

**Fig. 3.** $dU$ as a function of $f$ for (a) TeraFETs of Si and WBG materials (GaN and p-Diamond) under $\alpha = 0.3$ and (b) p-Diamond TeraFETs with $\alpha = -0.3 \sim 0.5$. $U_0 = -0.4$ V, $L = 130$ nm, $C_g = C_{g0}\exp(\alpha x/L)$.

Fig. 3(a) shows the frequency dependence of $dU$ under different $\alpha$ values for Si, AlGaN/GaN and p-Diamond (p-D) TeraFETs. The material parameters used are given in Table I. We can see that for p-D and AlGaN/GaN devices, the resonant peaks are observed, and thus the simulation results exhibit more significant deviation from the analytical curves compared to Si FET. Indeed, with larger effective masses and mobilities, the plasmonic resonance ($\omega\tau \geq 1$) can be achieved at much lower frequencies in wide-band-gap (WBG) materials compared to those in Si (see values of $\gamma/2\pi$ in Table I). Moreover, the p-D TeraFET exhibits the largest $dU$ within the sub-THz region, reflecting its advantage in detection sensitivity over other materials. This observation conforms to our previous results in [20] and [21]. Fig. 3(b) illustrates the $dU$ as a function of $f$ for p-D TeraFETs under different $\alpha$ values. For $f < 300$ GHz, $dU$ still increases with the rise of $\alpha$. Beyond 300 GHz, a much weaker $\alpha$ dependence is observed. Note that the fundamental resonance is reached at around 250 GHz, thus the modulation of $dU$ by $C_g$ at the resonant regime is achieved. At $f = 250$ GHz, $dU$ increases from 50.4 mV to 58.2 mV as $\alpha$ rises from 0 to 0.5, showing a 15% elevation. While for Si FETs, the increment is 12.9% (13.2 mV to 14.9 mV). Those data further demonstrated the potential advantages of p-D over other materials for plasmonic sub-THz detection applications.

**Table I**. Material parameters used in this work. (Note: $d_b$ and $\varepsilon_r$ are the thickness and permittivity of dielectric barrier layer, respectively. $\nu$ is the kinematic viscosity)

| Material | Si MOS | AlGaN/GaN | p-D |
|---|---|---|---|
| $m/m_0$ | 0.19 | 0.23 | 0.663 |
| $\mu$ (m$^2$/Vs) | 0.1 | 0.2 | 0.2 |
| $\gamma/2\pi$ (THz) | 1.47 | 0.61 | 0.21 |
| $d_b$ (nm) | 4 | 20 | 4 |
| $\varepsilon_r$ | 3.9 | 5.7 | 8.9 |
| $\nu$ (cm$^2$/s) | 5.6 | 4.6 | 1.6 |
| Reference | [50-52] | [53-55] | [21,56,57] |

To improve the tunability of $dU$ by $C_g$, it is intuitive to explore other $C_g$ profiles. Here we compare the detection performance of Si TeraFETs under 3 configurations (exponential, linear and sawtooth, see Fig. 1(b)). For non-exponential $C_g$ profiles, $\alpha_1$ is a function of $x$, and thus the term $n \cdot \partial \alpha_1 / \partial x$ becomes non-zero and contributes to the transport. This provides additional possibility for the $dU$ manipulation. Fig. 4(a) shows $dU$ as a function of $\alpha$ under different configurations. As seen, the $dU$ profile under linear configuration is

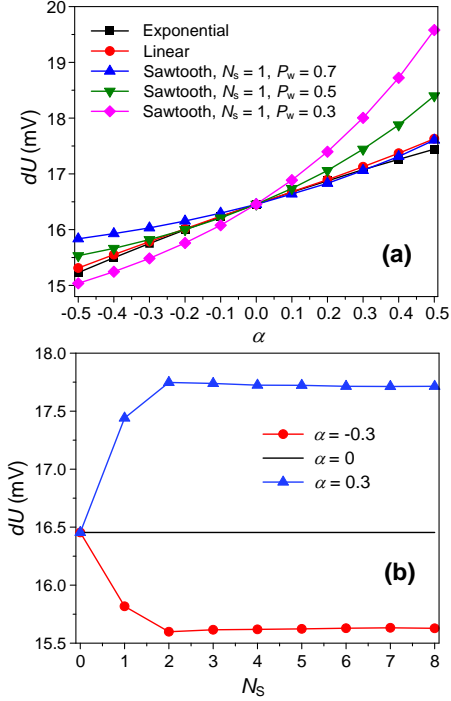

**Fig. 4.** Simulation results of $dU$ for Si TeraFETs (a) as a function of $\alpha$ under different $C_g$ configurations and (b) $dU$ as a function of $N_s$ with sawtooth profile under $P_w = 0.5$. $P_w$ is the ratio of rising edge period in each segment. $f = 350$ GHz, $U_0 = -0.4$ V.

similar to that of exponential configuration. However, the tunability of sawtooth configuration appears to be significantly better than other twos, as the response can be tuned by various parameters: $\alpha$, $P_w$ (the ratio of rising edge) and $N_s$ (number of segments). With the decrease of $P_w$, the tunability of $dU$ by $\alpha$ enhances. If $N_s$ increases, the $\alpha$-tunability also improves, as shown in Fig. 4(b), where the maximum tunability is observed at $N_s = 2$. Beyond $N_s = 2$, the simulated $dU$ saturates. The above response characteristics under different $C_g(x)$ profiles is related to the changes in $\alpha_1(x)$ as the parameters alter. Fig. 5 illustrates the spatial distribution of $\alpha_1(x)$ for linear, exponential, and sawtooth $C_g(x)$ configurations. Apparently, $\alpha_1$ curves of sawtooth profiles in Fig. 5(b) exhibit larger peak values and more violent oscillations compared to those of linear and exponential configurations shown in Fig. 5(a). This might be the reason why a sawtooth $C_g(x)$ has a better $dU$ tunability than other configurations. As $N_s$ and $P_w$ changes, the peaks and ratios of positive and negative regions in $\alpha_1(x)$ could change significantly,

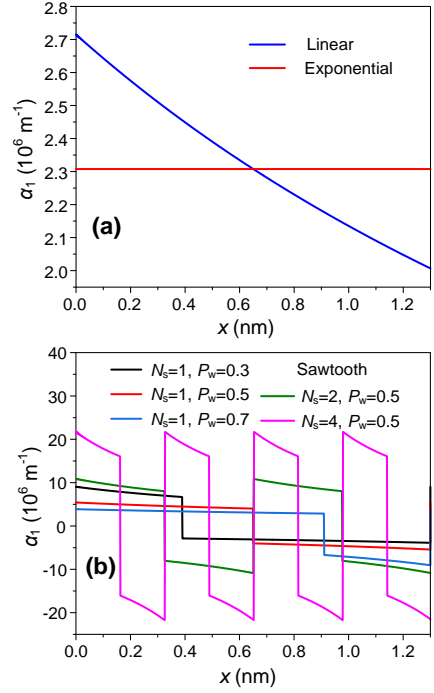

**Fig. 5.** The spatial distribution of $\alpha_1$ for (a) linear and exponential $C_g(x)$ profiles and (b) sawtooth profiles. Parameters: $L = 130$ nm, $\alpha = 0.3$.

as shown in Fig. 5(b), which further modifies the detection performance.

In addition to the non-uniform $C_g$ structures, the manipulation of DC response can also be achieved by a spatially non-uniform threshold voltage $V_{th}$ under a uniform $C_g$. The non-uniform $V_{th}$ can be achieved by the doping modulation[36,37]. To analytically evaluate the effects of non-uniform $V_{th}$, we consider the original form of Navier-Stocks equation for carriers[58]:

$$\frac{\partial \boldsymbol{v}}{\partial t} + (\boldsymbol{v} \cdot \nabla)\boldsymbol{v} + \gamma \boldsymbol{v} = \frac{e}{m}\nabla \varphi \quad (9)$$

Note that $U = V_g - V_{th} - \varphi$, for a uniform $V_{th}$, $\Delta\varphi = -\Delta U$ and (9) reduces to (3). With a non-uniform $V_{th}$, we obtain

$$\frac{\partial \boldsymbol{v}}{\partial t} + (\boldsymbol{v} \cdot \nabla)\boldsymbol{v} + \gamma \boldsymbol{v} + \frac{e}{m}(\nabla U + \nabla V_{th}) = 0 \quad (10)$$

and under the non-resonant operation (10) reduces to

$$\boldsymbol{v} = -\frac{e\tau}{m}(\nabla U + \nabla V_{th}) \quad (11)$$

Combining (11) and (2), we obtain (see Section D of Supplementary Material)

$$\frac{\partial n}{\partial t} - S_e^2\tau\frac{\partial^2 n}{\partial x^2} - S_0^2\tau\frac{\partial(\alpha_1^* n)}{\partial x} = 0 \quad (12)$$

Where we define $\alpha_1^* = \frac{1}{|V_{th0}|}\frac{\partial V_{th}}{\partial x}$, $S_0 = \sqrt{\frac{e|V_{th0}|}{m}}$.

Eq. (12) has a similar form as Eq. (5) and can be solved by the same method. Now the detection performance is controlled by $\alpha_1^*$. (12) and (5) becomes identical when

$$S_e^2 \alpha_1 + S_0^2 \alpha_1^* = 0 \qquad (13)$$

Eq. (13) indicates that $\alpha_1^*$ needs to be negative in order to improve the detection sensitivity. i.e. the threshold voltage should decrease from source to the drain so as to elevate $dU$.

In conclusion, we successfully manipulated the DC voltage response of TeraFETs by non-uniform capacitance profiles, and proposed a theory to qualitative explain this phenomena. The analytical calculations showed a good agreement with the simulation data under the non-resonant, subthreshold condition. In general, the DC response voltage is elevated if $C_g$ rises from source to drain. We compared the modulation performances of 3 different $C_g(x)$ profiles and found that a sawtooth configuration exhibited the largest response tunability. The manipulation of DC response can also be achieved by a spatially non-uniform threshold voltage. Those discoveries could help improve the responsivity of plasmonic TeraFET detectors and promote their industrial applications.

**SUPPLEMETARY MATERIAL**

See Supplementary Material for details of gate capacitance profiling, derivations of DC response voltage, and the effects of device loading and gate leakage current.

**DATA AVALIBILITY**

The data that support the findings of this study are available from the authors upon reasonable request.